\documentclass[usenatbib]{basi}
\usepackage[british]{babel}
\usepackage[varg]{txfonts}
\usepackage{rotating}
\usepackage{dcolumn}
\usepackage{aas}
\usepackage{url}

\begin{document}
\title{Using a model for telluric absorption in full-spectrum fits}
\author[T.-O.~Husser and K.~Ulbrich]%
       {T.-O.~Husser$^1$\thanks{email: \texttt{husser@astro.physik.uni-goettingen.de}} and K.~Ulbrich$^1$\\
       $^1$Institut f\"ur Astrophysik, Georg-August-Universit\"at G\"ottingen, Friedrich-Hund-Platz 1, \\ \quad 37077 G\"ottingen, Germany}

\pubyear{2014}
\volume{10}
\pagerange{\pageref{firstpage}--\pageref{lastpage}}

\date{Received \today}

\maketitle
\label{firstpage}

\begin{abstract}
The typical approach for removing telluric absorption lines from a science spectrum is to divide it
by the spectrum of a standard star of spectral type A or B observed close in time and airmass. We present
a new method, where we use a model for the transmission of the Earth's atmosphere in a full-spectrum fit,
which determines the parameters for the stellar and Earth's atmosphere simultaneously. This eliminates
the need of a standard star completely.
\end{abstract}


\section{Introduction}
The absorption lines caused by molecules in the Earth's atmosphere are a problem every astronomer faces 
when doing ground-based spectroscopy. While these so-called telluric lines can be found at
known positions, their strengths depends on the current weather conditions at the site of the
observatory.

Since the telluric lines often cover parts of the spectra that are of interest, one usually tries to
remove them. Most commonly this is done by observing early-type stars (spectral types mid-B or late-A)
that show only few and weak  metal lines. In order to remove the telluric lines from a science spectrum 
it is then divided by the spectrum of this standard star. Unfortunately, there are several problems with this method:
(a) a suitable telluric standard star is required that has been observed both close in time to the science 
observation and at a similar airmass, which (b) also requires more telescope time. Furthermore, (c) A/B type 
stars show strong hydrogen absorption features that need to be taken into account when removing the tellurics.
The latter item can be addressed by interpolating linearly over the hydrogen lines, but this
only works well if one is not particularly interested in these regions of the spectrum. In order to avoid
this, several attempts have been made to use high-resolution spectra of the Sun and Vega in combination
with standard stars of spectral types F/G \citep{1996AJ....111..537M} and A0 V \citep{2003PASP..115..389V} 
respectively.

\begin{figure}
  \includegraphics{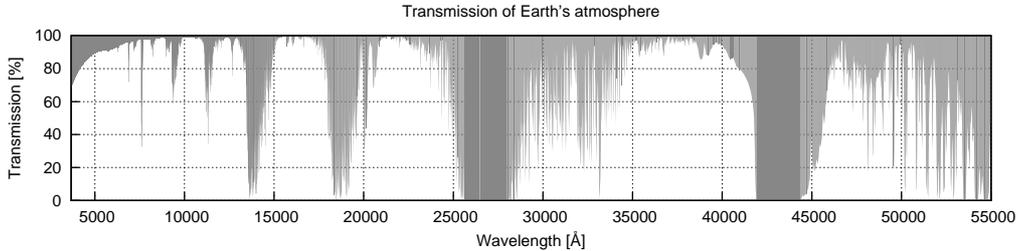}
  \caption{Synthetic transmission spectrum of the Earth's atmosphere.}
  \label{figure:tellurics}
\end{figure}
The need of a standard star can be avoided completely when using theoretical models for the atmospheric
transmission, which for this work have been computed using LBLRTM\footnote{\url{http://rtweb.aer.com/lblrtm_frame.html}}
(Line-By-Line Radiative Transfer Model), which in turn is based on FASCODE \citep{2005JQSRT..91..233C,1992JGR....9715761C}.
Since this has been used before for removing telluric lines by \cite{2010A&A...524A..11S} and \cite{rudolf}, we refer
to their publications for further details. A synthetic transmission spectrum created by this method is shown
in Fig.~\ref{figure:tellurics}.

\section{Fitting}
While \cite{2010A&A...524A..11S} and \cite{rudolf} fit the tellurics either with a fixed stellar spectrum or only
in those regions of the spectrum that are free of intrinsic stellar absorption features, we combined it with our methods for full-spectrum
fitting, which works similarly to, e.g., the ULySS package \citep{2009A&A...501.1269K}.

\begin{figure}[ht!]
  \includegraphics{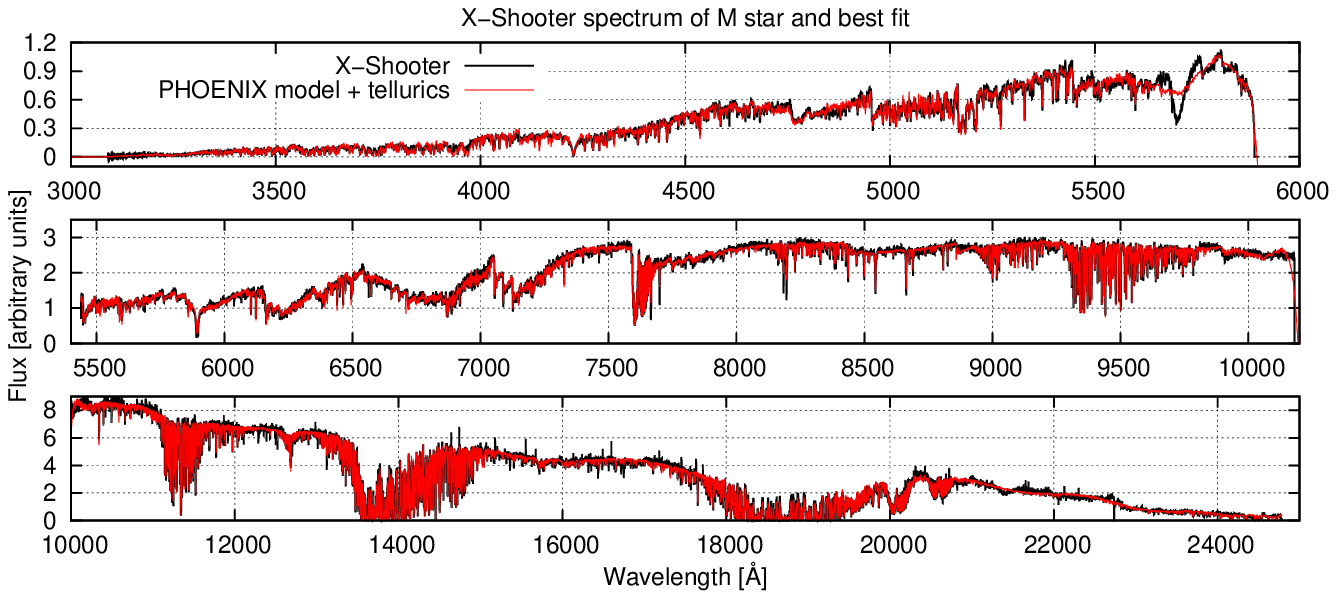} \\
  \includegraphics{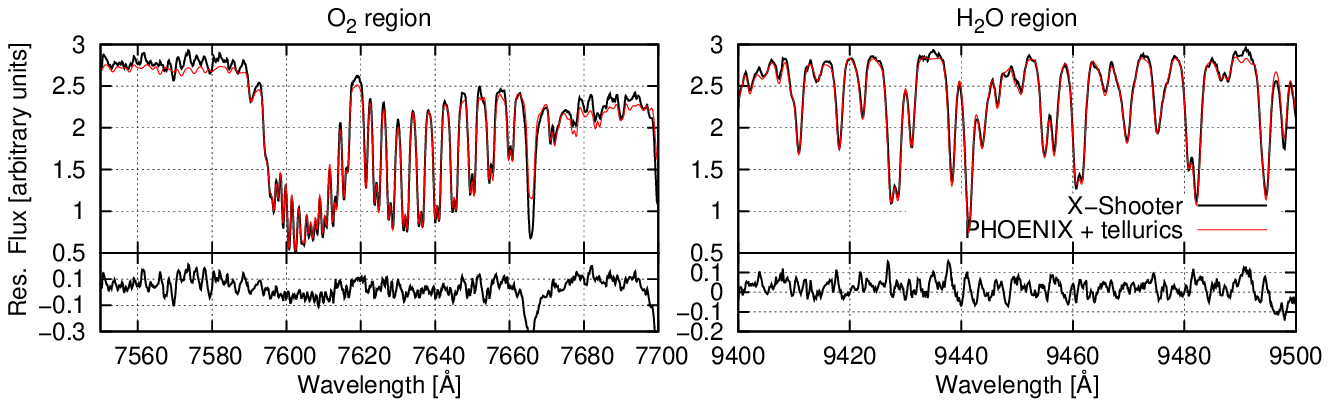}
  \caption{Full X-Shooter spectrum (by S.\ Sch\"afer, priv.\ comm.), and the best-fit combination of a PHOENIX spectrum and 
           a telluric model are shown in the upper three panels. Below zooms into a $O_2$ and a $H_2O$ region are shown with residuals.}
  
  \label{figure:xshooter}
\end{figure}
The fundamental problem in a full-spectrum fit is finding a model spectrum $G(x)$ that best describes an observation
$O(x)$ in pixel space. We describe the model as the weighted sum of single model spectra $M_k$ with coefficients $\omega_k$, 
convolved with corresponding kernels $B_k$:
\begin{equation}
 G = \sum\limits_{k=1}^{K} \omega_k \left[ B_k \ast M_k \right],
\end{equation}
where $\ast$ denotes the convolution in the logarithmic domain, which allows us to handle both line shifts (i.e.\ 
radial velocities) and broadenings in one step.

To allow for a telluric model to be included in the fit, we created small model grids with varying abundances for some of the
most prominent molecules in the Earth's atmosphere ($H_2O$, $CO_2$, $O_2$, $O_3$, $N_2O$, and $CH_4$). Since LBLRTM always includes water in
the atmospheres, we had to calculate the water-free models by dividing with the pure $H_2O$ model, which of course produces
some inaccuracies in water dominated regions of the spectra, but seems to be negligible. An interpolator then allows
us to create absorption spectra $T_n$ for each molecule with arbitrary abundances. The final telluric spectrum 
is just the product of all those spectra, convolved with a given kernel $C$ for degrading them to the instrumental profile
and applying a line shift:
\begin{equation}
 G = \sum\limits_{k=1}^{K} \omega_k \left[ B_k \ast M_k \right] \cdot C \ast \prod\limits_{n=1}^{N} T_n.
\end{equation}

\begin{figure}[ht!]
  \includegraphics{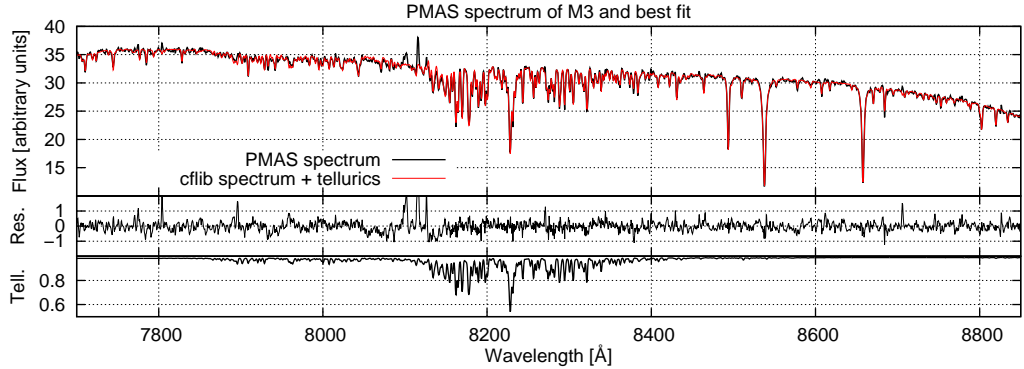}
  \caption{PMAS spectrum of star in M3 (by S.\ Kamann, priv.\ comm.) and best fit of cflib spectrum and 
           telluric model with residuals. The tellurics only are plotted at the bottom.}
  \label{figure:pmas}
\end{figure}
The additional term adds $N+2$ more free parameters to the fit, i.e.\ one for each of the $N$ molecules plus line shift
$\sigma$ and broadening $v$. 
Depending on the wavelength range, $N$ can be as small as one, e.g.\ when the spectrum is limited to the optical
domain ($H_2O$ only). Using a fast interpolation scheme for the tellurics, the computation time increases only slightly
compared to a fit without tellurics.

\section{Examples \& Conclusion}
In Fig.~\ref{figure:xshooter}, a full X-Shooter spectrum of an M star is shown together with the best fit of a PHOENIX spectrum
\citep[\emph{G\"ottingen Spectral Library by PHOENIX}\footnote{\url{http://phoenix.astro.physik.uni-goettingen.de/}}, see][]{2013A&A...553A...6H}
and a telluric model. Clearly, both match very well, especially in the regions dominated by telluric absorption features as
shown in the detail plots in the lower two panels. Another example is shown in Fig.~\ref{figure:pmas}, where a PMAS
spectrum of a star in M3 is shown together with a matching spectrum from \emph{cflib} \citep{2004ApJS..152..251V} combined
with a telluric model. Again, we achieved a very good agreement.

In this paper, we introduced a very simple method for integrating a synthetic model for the transmission of the Earth's atmosphere
into a full-spectrum fit. We showed that it performs well and allows us to eliminate the need of observing a telluric standard 
star completely.


\label{lastpage}
\end{document}